\pdfoutput=1
\documentclass[aps,pre,twocolumn,superscriptaddress,groupedaddress]{revtex4}

\newcommand{\arcsec}{\ifmmode^{\prime\prime}\else $^{\prime\prime}$\fi}
\newcommand{\arcmin}{\ifmmode^{\prime}\else $^{\prime}$\fi}
\newcommand{\degrees}{\ifmmode^{\circ}\else $^{\circ}$\fi}

 \newcommand{\bfn}{{\bm n}}

\newcommand{\bfu}{{\bm u}}

\newcommand{\bfB}{{\bm B}}

\newcommand{\bfP}{{\bm P}} \newcommand{\bfQ}{{\bm Q}}

\newcommand{\Rm}{{R_{\rm m}}}

\usepackage{amssymb, amsmath}
\usepackage{graphicx}
\usepackage{bm}
\usepackage{url}
\usepackage{color}
\begin{document}

\title{Growth rate degeneracies in kinematic dynamos}%

\author{B. Favier}%
\email[Corresponding author: ]{b.favier@damtp.cam.ac.uk}
\author{M.R.E. Proctor}
\affiliation{Department of Applied Mathematics and Theoretical Physics, University of Cambridge, Centre for Mathematical Sciences, Wilberforce Road, Cambridge CB3 0WA, UK}
\date{\today}

\begin{abstract}
We consider the classical problem of kinematic dynamo action in simple steady flows.
Due to the adjointness of the induction operator, we show that the growth rate of the dynamo will be exactly the same for two types of magnetic boundary conditions: the magnetic field can be normal (infinite magnetic permeability, also called pseudo-vacuum) or tangent (perfect electrical conductor) to the boundaries of the domain.
These boundary conditions correspond to well-defined physical limits often used in numerical models and relevant to laboratory experiments.
The only constraint is for the velocity field $\bm{u}$ to be reversible, meaning there exists a transformation changing $\bm{u}$ into $-\bm{u}$.
We illustrate this surprising property using $S_2T_2$ type of flows in spherical geometry inspired by \cite{dudley89}.
Using both types of boundary conditions, it is shown that the growth rates of the dynamos are identical, although the corresponding magnetic eigenmodes are drastically different. 
\end{abstract}

\maketitle

%
%#####################################################################################################
%
%\section{Introduction}

The growth of magnetic fields due to dynamo action, both in astrophysical bodies and in laboratory experiments,  is expected to depend not only on the details of the flow field, but also on the conditions on the magnetic field applied at the boundaries.
In the laboratory there are two physically important limits: perfectly conducting, implying no normal field; and normal field, otherwise infinite permeability, where the tangential field at the boundary is zero.
These conditions are so different that one might expect that the dynamo properties would be quite different in the two cases.
In general this is true, but there is an important class of flows for which this is not the case. We call these {\it reversible} flows, defined as follows: consider the group $D$ of transformations which leave the boundaries invariant; then a velocity field $\bm u({\bm x})$ is reversible if ${\bm u}({\bm x})=-{\bm u}({\sf d}\cdot {\bm x})$, for some ${\sf d} \in D$. In other words, one can reverse the direction of the flow by an appropriate transformation. Then the main result of this paper can then be stated as follows:

Consider a steady flow of an electrically-conducting fluid of constant magnetic diffusivity $\eta$, contained in a volume $V$ and delimited by boundaries $S$.
\textit{Providing that the velocity field is reversible in the above sense, the growth rate of the kinematic dynamo will be exactly the same whether the boundaries are made of a perfect electrical conductor or have an infinite magnetic permeability. In fact the whole spectrum of growth rates will be identical.}
This remarkable result is due to the adjointness property of the induction operator as discussed by \cite{roberts60,gibrob67,proctor77a,proctor77b}. It should be noted that there is no statement about the relation between the respective eigenfunctions and indeed as seen below these might differ considerably in the two cases.

This result is formally proved as follows:
We begin with an eigenfunction for the growing magnetic field $\bfB$ satisfying the perfectly-conducting boundary condition $\bm{B}\cdot\bm{n}=0$, where $\bm{n}$ is the unit vector normal to the surface $S$.
The electric field must be normal to the boundaries and the tangential electric current vanishes there, $\left(\nabla\times\bm{B}\right)\times\bm{n}=0$ on $S$.
The equation for the magnetic potential can be written using the Weyl gauge as
\begin{equation}
\label{eq:induceigen}
s\bm{A}=\bfu\times\bfB-\eta\nabla\times\bfB \ ,
\end{equation}
where $s$ is the complex growth rate and $\bm{A}$ is the magnetic vector potential defined by $\nabla\times\bm{A}=\bm{B}$.
Since we have $\bm{u}\cdot\bm{n}=\bm{B}\cdot\bm{n}=\left(\nabla\times\bm{B}\right)\times\bm{n}=0$ on the boundary $S$, the cross product of equation \eqref{eq:induceigen} with $\bm{n}$ implies that $\bm{A}\times\bm{n}=0$ on $S$.
Then if we multiply the complex conjugate of equation \eqref{eq:induceigen} by a solenoidal vector field $\bfQ=\nabla\times\bm{P}$, and integrate over the entire domain $V$, we obtain after integrating by parts
\begin{multline}
\label{eq:adjoint}
\int_V\bfB^*\cdot(s^*\bfP+\bfu\times\bfQ+\eta\nabla\times\bfQ) \ \textrm{d}V= \\ - s^*\int_S\left(\bm{P}\times\bm{A}^*\right)\cdot\bm{n} \ \textrm{d}S - \eta\int_S\left(\bm{B}^*\times\bm{Q}\right)\cdot\bm{n} \ \textrm{d}S \ ,
\end{multline}
where $s^*$ is the complex conjugate of $s$.
The first surface integral on the right-hand side of equation \eqref{eq:adjoint} vanishes since $\bm{A}\times\bfn=0$ at the boundaries.
The second surface integral vanishes providing that we specify $\bfQ\times\bfn=0$ at the boundaries.
This last condition trivially implies that the normal electric current associated with $\bm{Q}$ vanishes on $S$, \textit{i.e.} $\left(\nabla\times\bm{Q}\right)\cdot\bm{n}=0$.
The expression in parentheses on the left-hand side of equation \eqref{eq:adjoint} is then the operator on $\bfP$ adjoint to the original operator \eqref{eq:induceigen}.
Assuming that the eigenvectors $\bm{B}$ form a complete set, and taking the curl of this expression we obtain the following equation
\begin{equation}
s^*\bfQ=-\nabla\times(\bfu\times\bfQ)-\eta\nabla\times\nabla\times\bfQ \ ,
\end{equation}
which is the induction equation for the solenoidal vector $\bfQ$ with $\bfu$ replaced by $-\bfu$; now however $\bfQ$ satisfies the infinite magnetic permeability condition $\bfQ\times\bfn=\left(\nabla\times\bm{Q}\right)\cdot\bm{n}=0$ at the boundaries.
This shows that interchanging the boundary conditions and reversing the direction of the velocity field gives the same spectrum.
%In the present case however the velocity field has the property that $\bfu$ and $-\bfu$ are related by a simple reflectional symmetry since they are eigenvalues of the stability problem.
In consequence the growth rates as a function of the magnetic Reynolds number $\Rm$ will be the same for both sets of boundary conditions.
Note that the change in the direction of the velocity field for the adjoint problem has been known for a long time \citep{roberts60}.
The problem was to find the appropriate choice of boundary conditions for both the original and the adjoint problem \citep{proctor77b}.
Since most of the studies were motivated by the geodynamo problem, the external boundary condition for the original problem corresponded to a vacuum.
In that case, the general boundary condition for the adjoint problem is unknown apart from some particular cases \citep{gibrob67}.
The present demonstration shows that the adjoint boundary conditions associated with a perfect electrical conductor is an infinite magnetic permeability, both of them corresponding to clear physical limits.

%
%#####################################################################################################

%
\begin{figure}
      \includegraphics[width=70mm]{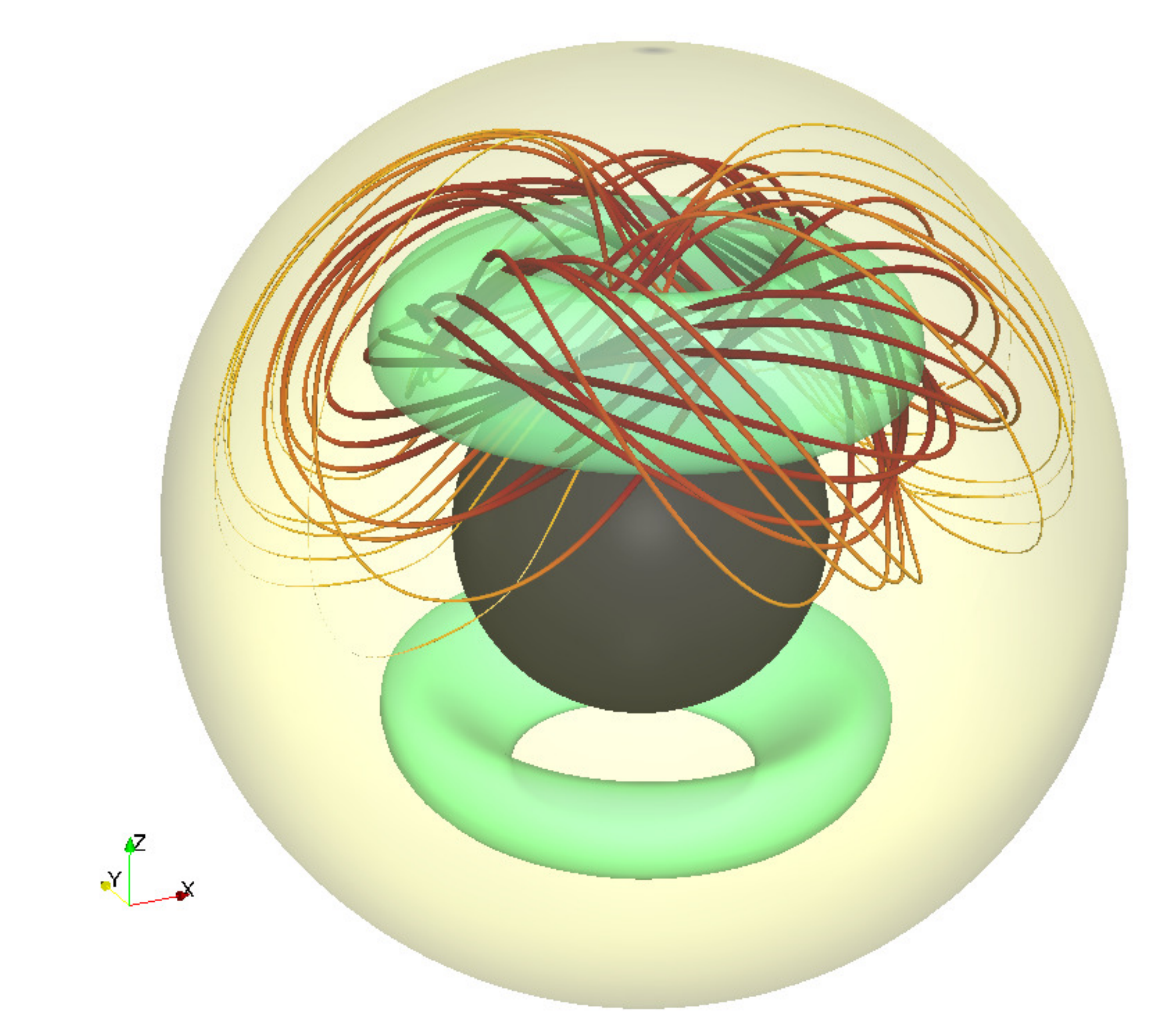}
      \includegraphics[width=70mm]{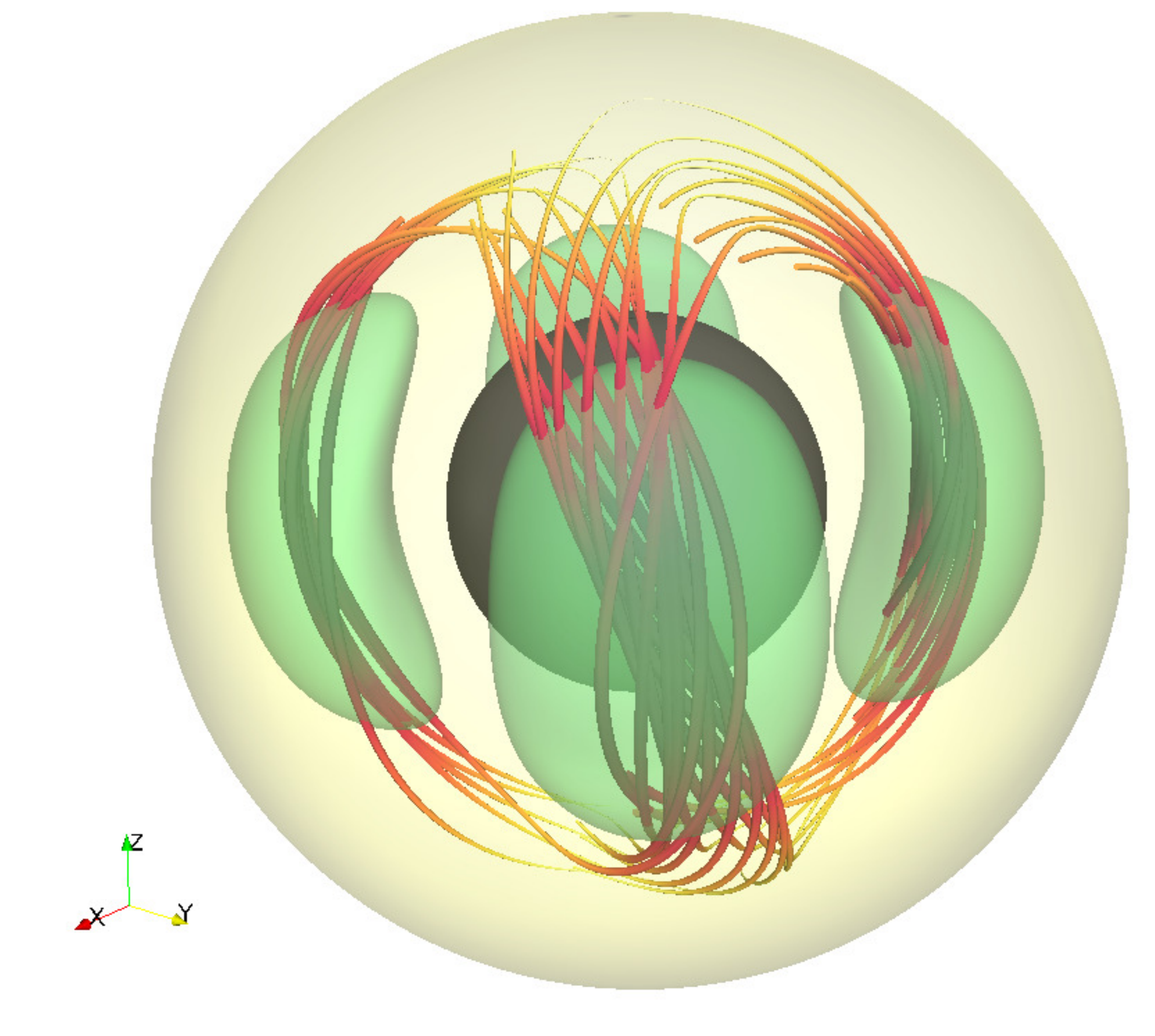}
    \begin{picture}(10,0)
        \put(-15,22){\large{$S_2^2T_2^2$}}
        \put(-15,202){\large{$S_2^0T_2^0$}}
    \end{picture}
    \caption{(Color online) Illustration of the $S_2T_2$ velocity fields considered in this paper. These flows are characterised by an azimuthal wavenumber $m=0$ (top) or $m=2$ (bottom) and a Legendre polynomial order of $l=2$. The aspect ratio is $\alpha=0.4$. The isosurfaces show the velocity magnitude at 75\% of its maximum value. The streamlines are randomly initiated inside one of the hemisphere. The dark and thick streamlines correspond to large velocity magnitude whereas bright and thin strealines correspond to low velocity magnitude. The axisymmetric flow on the top is \textit{not} reversible whereas the flow on the bottom is.}
    \label{fig:vel}
\end{figure}
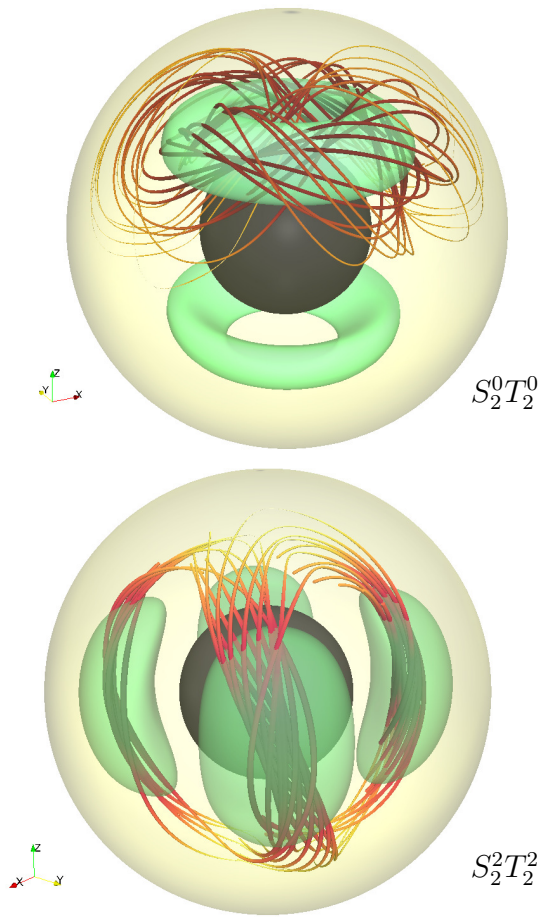
%
%#####################################################################################################
%
%\section{Illustration in spherical geometry\label{sec:spherical}}
%

We now illustrate our result in spherical coordinates $\left(r,\theta,\phi\right)$.
The choice of the coordinate system is not important and one could equally choose Cartesian or cylindrical coordinates.
We focus here on the spherical case as the differences in the magnetic eigenmodes when varying the boundary conditions are the most striking.
We consider an incompressible flow in a spherical shell defined by $\alpha<r<1$.
Kinematic dynamos driven by simple flows are a classical problem in dynamo theory and a lot of examples have been considered in the past (see for example \cite{dudley89} and references therein for the case of a full sphere).
The objective is here to compare kinematic dynamo action in two different flows with the two different types of boundary conditions mentioned previously.
The velocity field is first written using a poloidal-toroidal decomposition, thus ensuring incompressibility,
\begin{equation}
\bm{u}=\nabla\times\nabla\times\left(S\bm{e}_r\right)+\nabla\times\left(T\bm{e}_r\right) \ ,
\end{equation}
where $T$ is the toroidal component whereas $S$ is the poloidal component, and $\bm{e}_r$ is the unit vector in the radial direction.
Each of these scalars is then projected onto spherical harmonics, for example for the poloidal component,
\begin{equation}
\label{eq:sph}
S=\sum S_l^m(r)Y_l^m(\theta,\phi) \ ,
\end{equation}
where the sum is carried over integers such that $l \le m \le 0$, and $Y_l^m(\theta,\phi)$ is the classical spherical harmonic of azimuthal wave number $m$ and Legendre function order $l$.
The flows we consider in this letter are defined as follows: all coefficients $S_l^m(r)$ and $T_l^m(r)$ are zero except the ones for which $l=2$.
This type of flow is often referred as to a $S_2T_2$ flow.
In the azimuthal direction, all coefficients are zero except for one particular azimuthal wave number $M$ for which we impose
\begin{equation}
\label{eq:polvel}
S_2^M(r)=\sin^2\left(\pi\frac{r-\alpha}{1-\alpha}\right) \ ,
\end{equation}
\begin{equation}
\label{eq:torvel}
T_2^M(r)=8\sin^2\left(\pi\frac{r-\alpha}{1-\alpha}\right) \ .
\end{equation}
The factor $8$ in equation \eqref{eq:torvel} is arbitrarily introduced to minimise the critical magnetic Reynolds number for dynamo action to occur.
This choice of radial structure is compatible with an impenetrable ($S=0$) and no-slip ($T=\partial S/\partial r=0$) boundary condition for the velocity field.
We consider two possibilities for the azimuthal dependence: $M=0$ and $M=2$.
These two flows are naturally labelled $S_2^0T_2^0$ and $S_2^2T_2^2$ respectively.

The first flow has been studied in details in various geometries since it is a simple model of the mean-flow in the VKS experiment \citep{monchaux2007, gissinger2009}.
The flow corresponds to two axisymmetric helical cells in each hemisphere with net helicity throughout the domain, \textit{i.e.} $\mathcal{H}=\int_V\bm{u}\cdot\nabla\times\bm{u}\textrm{d}V\neq0$.
Note however that our conclusion does not depend on the presence or not of net kinetic helicity in the system.
This flow is \textit{not} reversible as defined earlier.
%There are two poloidal eddies with inward flow in the midplane and outward flow at the equator, which is not not symmetric if the direction of $\bm{u}$ is opposite.
An illustration of this steady flow can be found in figure \ref{fig:vel}.
Due to the axisymmetry of the flow, Cowling's theorem \citep{cowling33} forbids growing axisymmetric magnetic fields and the different azimuthal wave numbers of the magnetic field are decoupled.

The second flow is similar to the flow first studied by \cite{pekeris73} in the geodynamo context, albeit there is no inner core in their case.
It corresponds to a four cells flow with net kinetic helicity.
Due to its symmetries, this flow is reversible (a rotation of $\pi/2$ around the vertical axis changes $\bm{u}$ in $-\bm{u}$) and a visualisation for the particular aspect ratio $\alpha=0.4$ is shown in figure \ref{fig:vel}.
This simple type of flows is known to be a very efficient kinematic dynamo without an inner core \citep{dudley89}.

\begin{figure}
      \includegraphics[width=70mm]{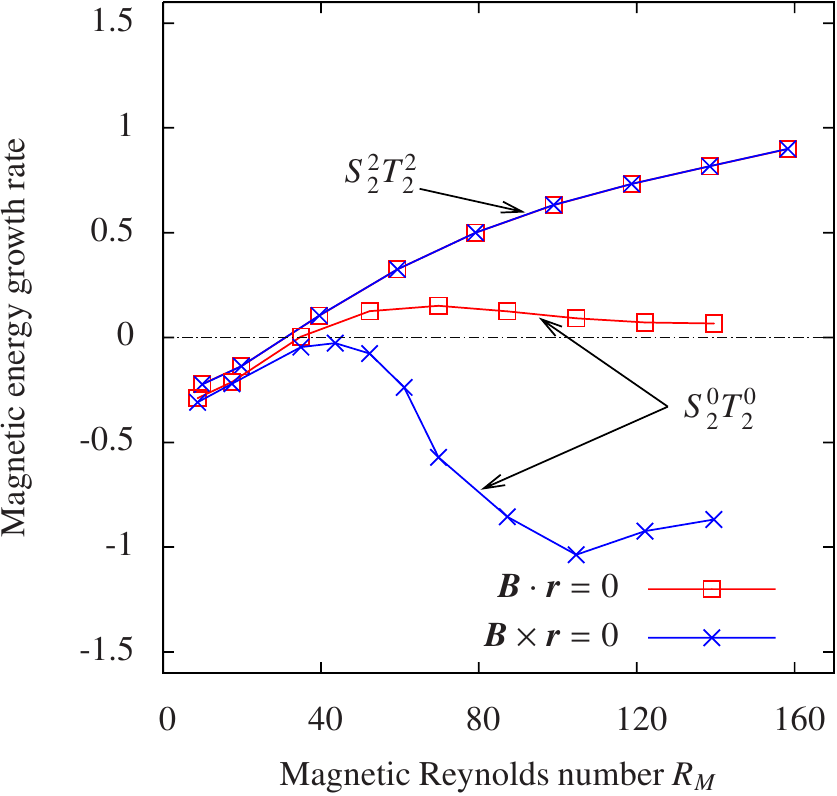}
    \caption{(Color online) Growth rate of the magnetic energy versus magnetic Reynolds number in the case of homogeneous boundary conditions. The square symbols correspond to the perfectly-conducting case where $\bm{B}\cdot\bm{r}=0$ at both boundaries whereas the cross symbols correspond to the perfectly-insulating case where $\bm{B}\times\bm{r}=0$. The results are shown for the aspect ratio $\alpha=0.4$. For the $S_2^0T_2^0$ flow, only the growth rates associated with the $m=1$ mode are shown.}
    \label{fig:growth}
\end{figure}
\begin{figure*}
      \includegraphics[width=70mm]{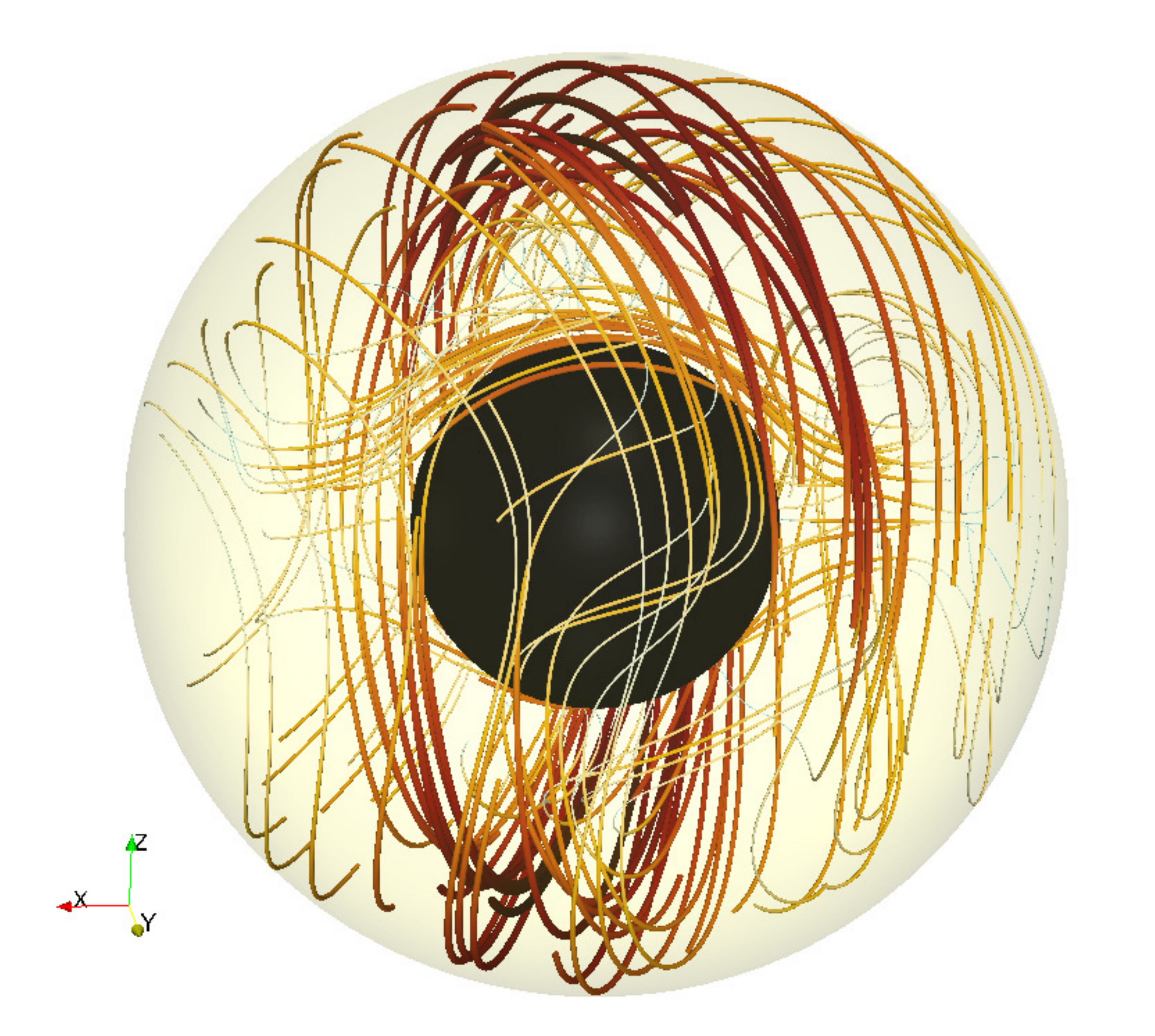}
      \includegraphics[width=70mm]{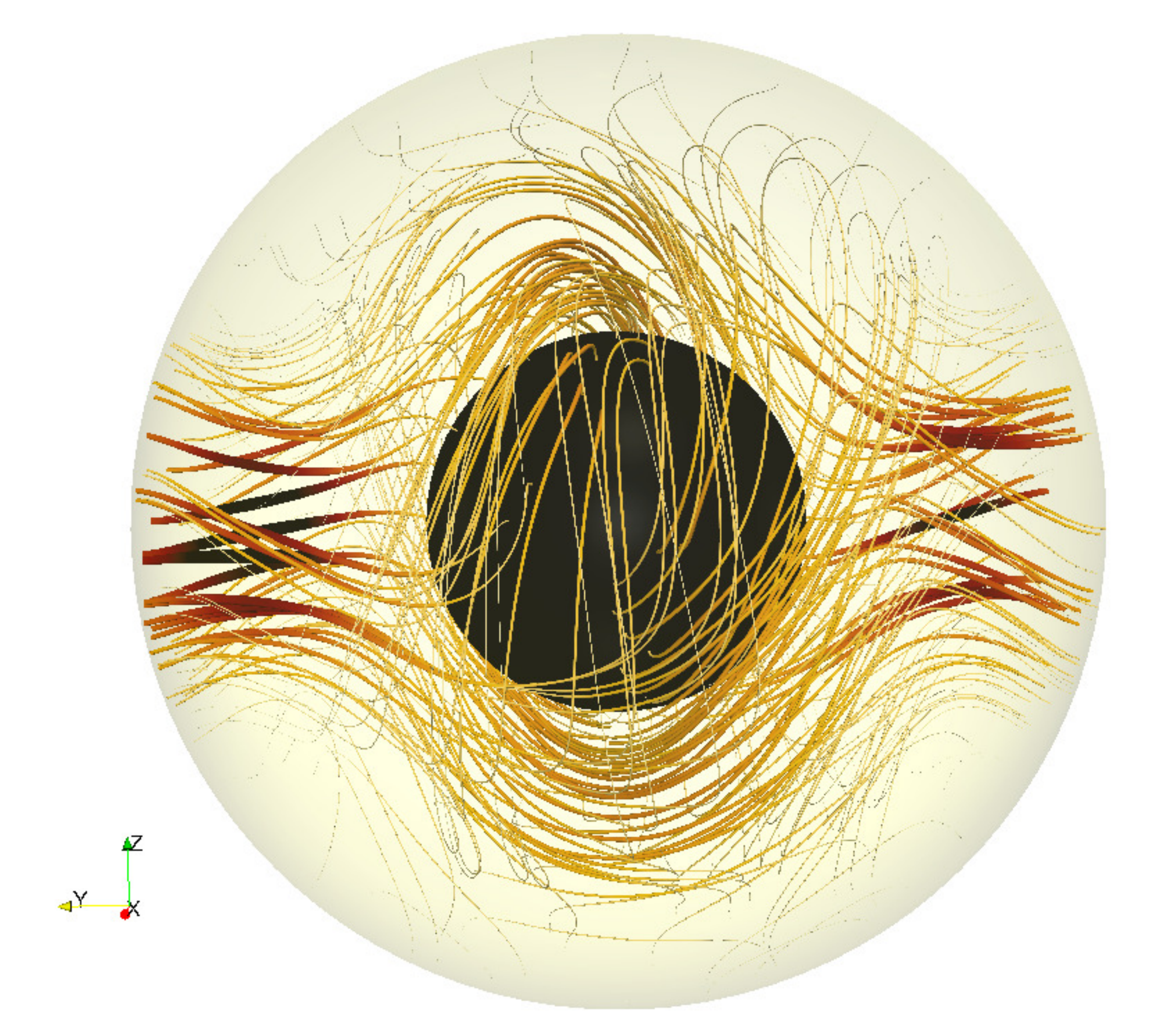} 
    \caption{(Color online) Magnetic eigenmodes close to the onset for kinematic dynamo action driven by the $S_2^2T_2^2$ reversible flow. Left: the boundary conditions are perfectly-conducting (no normal field). Right: the boundary conditions correspond to a pseudo-vacuum (no tangent field). In both cases, the magnetic field is dominated by a strong $m=1$ mode. The magnetic field lines are initiated randomly in the spherical shell. The dark and thick magnetic field lines correspond to large magnetic field amplitude whereas bright and thin lines correspond to low magnetic field magnitude. The growth rate associated with these two eigenmodes is exactly the same.}
    \label{fig:eigen}
\end{figure*}

In order to check our finding concerning the growth rate of kinematic dynamo action and its dependence on magnetic boundary conditions, we need to solve the induction equation with a prescribed velocity field.
While this problem is linear and could be reduced to an eigenvalue problem, the relatively large three-dimensional resolution required here to solve the induction equation makes the equivalent initial value problem much easier to handle.
%An eigenvalue problem would indeed require the inversion of a very large matrix.
As a consequence, the induction equation is solved using the numerical code PARODY.
This code was originally written by E. Dormy \citep{dormy1998} and later improved by J. Aubert \citep{aubert2008}.
PARODY has been benchmarked against other numerical codes in the context of a convectively-driven dynamo problem \citep{christensen2001}.
%It has then been used to consider various models, mostly related to the geodynamo.
Although the code is able to solve the full set of magnetohydrodynamics equations in the Boussinesq approximation, we only use the induction equation solver throughout this paper.
The solenoidal magnetic field is written using a poloidal and toroidal decomposition and both poloidal and toroidal scalars are then projected onto spherical harmonics, as in equation \eqref{eq:sph}.
The radial functions ${B_t}_l^m(r)$ for the toroidal field and ${B_p}_l^m(r)$ for the poloidal field are represented by their discretized values on a non-uniform radial grid.
The grid is denser close to the inner and outer boundaries in order to accurately resolve boundary effects.
The radial derivatives are computed using second order finite-differences.
In the case of a perfectly-conducting boundary condition, the poloidal and toroidal components of the magnetic field must verify the following constraint for all $l,m$
\begin{align}
\frac{\partial^2 B_p}{\partial r^2}+\frac{2}{r}\frac{\partial B_p}{\partial r} & = 0 \ , \\
\frac{\partial B_t}{\partial r}+\frac{1}{r}B_t & = 0 \ .
\end{align}
Note that due to the solenoidality of the magnetic field, these conditions directly imply that $B_p=0$ at the boundaries.
In the case of an infinite magnetic permeability, the corresponding boundary conditions are
\begin{align}
\frac{\partial B_p}{\partial r}+\frac{1}{r}B_p & = 0 \ , \\
B_t & = 0 \ .
\end{align}
The time-stepping is achieved using a semi-implicit Crank-Nicholson scheme for the diffusive term and a second order Adams-Bashforth scheme for the advective term.
The typical resolution is $480$ points in the radial direction, and a spherical harmonic decomposition truncated at $l,m<64$.
In the case of the $S_2^0T_2^0$ flow, since all azimuthal magnetic modes are decoupled, only the most unstable $m=1$ mode is considered.

We first compute the growth rate of the magnetic energy varying the magnetic Reynolds number defined here as
\begin{equation}
R_M=\frac{U_{\textrm{max}}(1-\alpha)}{\eta} \ ,
\end{equation}
where $U_{\textrm{max}}$ is the maximum velocity in the spherical shell.
We here consider a particular aspect ratio of $\alpha=0.4$ but our results do not qualitatively depend on this particular choice.
For the flows defined by equations \eqref{eq:polvel} and \eqref{eq:torvel}, we have $U_{\textrm{max}}=32.98$ for $M=2$ and $U_{\textrm{max}}=29.07$ for $M=0$.
The induction equation is then solved from an initial magnetic seed.
After a rapid transient phase during which the initial condition is forgotten, the magnetic energy is exponentially growing or decaying.
We compare in figure \ref{fig:growth} the results obtained varying the boundary conditions from perfectly-conducting to perfectly-insulating on both boundaries and for the two different flows.
As expected from the previous demonstration, the kinematic growth rates do not depend on the boundary conditions for the $S_2^2T_2^2$ flow.
The critical magnetic Reynolds number is approximately $R_M\approx40$ in this case.
The fact that the growth rates are exactly equal for both types of boundary conditions is even more surprising looking at the corresponding magnetic eigenmodes.
We show in figure \ref{fig:eigen} an illustration of the magnetic eigenmodes close to the onset of dynamo action.
As expected due to the effect of the boundary conditions, the magnetic topology is significantly different in both cases.
%The location of the maximum of magnetic energy is also varying significantly between the two types of boundary conditions.
The growth rate associated with these two eigenmodes is however exactly the same.

The growth rates for the two types of boundary conditions are however clearly distinct for the non-reversible $S_2^0T_2^0$ flow (see figure \ref{fig:growth}).
Since the azimuthal magnetic modes are decoupled, we only show the growth rates associated with the most unstable mode $m=1$.
In that case, a dynamo is observed in the case of perfectly-conducting boundary conditions whereas no dynamo at all is found with an infinite magnetic permeability.
As already mentioned, this flow shares some similarity with the mean velocity field of the VKS experiment.
The effect of the magnetic boundary conditions on the dynamo threshold of von K\'arm\'an swirling flows has been studied by \cite{gissinger2008}.
The lack of dynamo in the infinite magnetic permeability case is due to the presence of the large inner core in our case.
As the size of the core is reduced, we recover the dynamo observed by several studies, with a strong equatorial dipole.

We also considered different flows corresponding to different spherical harmonics, radial structures and spherical shell aspect ratios, and the conclusion remains qualitatively the same.
The previous result is also valid in the case of different boundary conditions at each boundary.
If the inner core is perfectly-conducting whereas the outer core is perfectly-insulating, the growth rate of the kinematic dynamo will be the same if we reverse the boundary conditions configuration and the direction of the reversible flow.

Finally, we considered different types of flows in different geometries.
For example, one can consider the flow resulting from rotating convection in the Boussinesq approximation just above onset.
In that case, the resulting steady flow in a plane layer model can correspond to square or hexagonal patterns \citep{veronis59}, which are all reversible.
We solved the induction equation for both patterns and also found that the eigenvalue spectrum is the same when varying the boundary conditions from a perfect conductor to an infinite magnetic permeability.
More details about kinematic dynamo action in such flows and the effect of boundary conditions can be found in \cite{favier2013pre}.
Note also that we have only discussed steady velocity fields up to now.
However, it seems that this result also holds for time periodic flows as long as the reversibility condition is valid at all times.
So far, we have only checked this result numerically by allowing the amplitude of the flow to be time dependent (not shown here) but a more general demonstration should be accessible.

%#####################################################################################################
%
%\section{Conclusion}
%

To conclude, we show in this letter that providing that a flow is reversible (as defined at the beginning of this letter), kinematic dynamo action will be the same with two different types of boundary conditions: the boundary can be either perfectly conducting, so that magnetic field lines are tangent to the surface, or can be of infinite magnetic permeability, so that magnetic field lines reconnect perpendicularly to the surface.
We verified this observation in spherical and Cartesian geometries for various types of flows.
While there is only a simple constraint on the velocity field for this result to be true, the required symmetry is however unlikely to be verified in a more realistic turbulent context.
It would therefore be interesting to consider the departure from this exact result in the experimentally relevant situation where small-scale velocity fluctuations are not reversible whereas the mean flow is.

\acknowledgments{The authors would like to thank Emmanuel Dormy and Toby S. Wood for valuable comments and suggestions. BF thanks the Cambridge Newton Trust for financial support.}

\bibliographystyle{unsrt}
\bibliography{biblio}

\end{document}